\RequirePackage{ifpdf}
\ifpdf 
\documentclass[pdftex]{sigma}
\else
\documentclass{sigma}
\fi

\usepackage{epic}
\usepackage{eepic}
\usepackage{young}
\usepackage{youngtab}

\newcommand{\yn}{\mathbb{Y}}
\newcommand{\yc}{\Yvcentermath1}
\newcommand{\mb}{\mbox{}}

\begin{document}

\allowdisplaybreaks

\renewcommand{\thefootnote}{$\star$}

\renewcommand{\PaperNumber}{024}

\FirstPageHeading

\ShortArticleName{Ordering of Energy Levels for Extended $SU(N)$ Hubbard Chain}

\ArticleName{Ordering of Energy Levels\\ for Extended $\boldsymbol{SU(N)}$ Hubbard Chain\footnote{This paper is a contribution to the Proceedings of
the XVIIIth International Colloquium on Integrable Systems and Quantum
Symmetries (June 18--20, 2009, Prague, Czech Republic).  The full
collection is
available at
\href{http://www.emis.de/journals/SIGMA/ISQS2009.html}{http://www.emis.de/journals/SIGMA/ISQS2009.html}}}

\Author{Tigran HAKOBYAN}

\AuthorNameForHeading{T. Hakobyan}

\Address{Yerevan State University, Alex Manoogian 1, Yerevan, Armenia\\
Yerevan Physics Institute, Alikhanian Br. 2, Yerevan, Armenia}

\Email{\href{mailto:hakob@yerphi.am}{hakob@yerphi.am}}

\ArticleDates{Received November 03, 2009, in f\/inal form March 15, 2010;  Published online March 20, 2010}

\Abstract{The Lieb--Mattis theorem on the antiferromagnetic ordering of energy levels
is generalized to $SU(N)$ extended Hubbard model with Heisenberg exchange and
pair-hopping terms.
It is proved that the minimum energy levels among the states from equivalent
representations are nondegenerate and ordered according to the dominance order
of corresponding Young diagrams. In particular, the ground states form a unique
antisymmetric multiplet. The relation with the similar ordering
among the spatial wavefunctions  with dif\/ferent symmetry  classes
of ordinary quantum mechanics  is discussed also.}


\Classification{81R05; 82D40; 82B20}

\section{Introduction}

The investigation of the ground state of quantum many-body systems is
very important and relevant in many areas of condensed matter physics,
in particular,  in high-temperature superconductivity and magnetism.
The quantum numbers associated with dif\/ferent symmetries and the degree of degeneracy
are essential properties that characterize the ground state
and the system as a whole.

In 1955, Marshall proved that the ground state of
antiferromagnetic Heisenberg spin-$1/2$ ring with an even number of
sites  is a spin singlet~\cite{M55}. This result
was generalized to higher spins and dimensions~\cite{LSM61}. The
uniqueness of ground state was established also. Moreover, for
Heisenberg antiferromagnets on bipartite lattices, Lieb and Mattis
proved that the ground state is a unique multiplet of spin
$S_\text{gs}=|S_1-S_2|$, where $S_1$, $S_2$
are the highest possible spin values of the two sublattices, which form
the bipartite lattice. They proved also that the lowest energy~$E(S)$ among all spin-$S$ states is an increasing function for
$S\ge S_\text{gs}$~\cite{LM62,Lieb89}. This property of the spectrum is known as
the antiferromagnetic ordering of the energy levels.
In one dimension, the  quantum mechanical system of interacting  electrons
without velocity- or spin-dependent forces
and the Hubbard model  possess this type of ordering  too \cite{LM62'}.
This fact is known as the absence of one-dimensional ferromagnetism.

The Lieb--Mattis theorem has been subsequently generalized to various spin
and fermion lattice systems, such as the spin-1 chain with
biquadratic interactions \cite{murno76}, the $t-J$~\cite{Amb92-1} and
extended Hubbard \cite{Amb92-2} chains, Hubbard models on bipartite
lattices at half f\/illing \cite{Shen98}, the Sutherland chain~\cite{H04},
frustrated spin-1/2 ladder systems~\cite{H07}. The ferromagnetic
ordering of the energy levels for spin-$1/2$ Heisenberg chain have
been established also~\cite{Nach03}.

In this article we formulate and prove the analogue of this theorem
for $SU(N)$ symmetric extended Hubbard--Heisenberg chains with pair-hopping
term.
Recently, the spin and fermionic chains with such symmetry
have been investigated intensively \cite{H04,Greiter07,Rachel09,Aguado09}.
This interest is motivated by their application in ultracold  atoms
\cite{Honercamp04,Gorshkov09}.
Recently, an interesting classif\/ication of the low-energy behavior of $SU(N)$
spin chains on three dif\/ferent types, which extends the well-known
classes of integer and half-integer spin chains \cite{Haldane83},
has been  conjectured and checked~\mbox{\cite{Greiter07,Rachel09}}.

In Section~\ref{section2} we introduce the model and describe the symmetries.
In Section~\ref{section3} we construct a basis, in which all of\/f-diagonal elements
of the Hamiltonian are non-positive. Section~\ref{section4} is devoted to the proof
of the uniqueness of the lowest energy states in the weight subspaces of $SU(N)$ algebra.
Here we also f\/ind the quantum numbers of such states.
Based on the results of previous sections, in Section~\ref{section5} we formulate and prove the analogue
of the antiferromagnetic ordering of energy levels for the  system under
consideration.  Its consequence for the ground state and its quantum numbers is analyzed in
Section~\ref{section6}. In the last section we establish the similar  ordering rule for the quantum
mechanical system of fermions with~$SU(N)$
degrees of freedom, and discuss the relation with~$SU(N)$ Hubbard model.

\section[$SU(N)$ symmetric fermionic chain]{$\boldsymbol{SU(N)}$ symmetric fermionic chain}\label{section2}

Consider the f\/inite-size $SU(N)$ symmetric extended Hubbard chain described by the Hamiltonian
\begin{gather}
H=\sum_{x,\alpha} -t_x\big( c^+_{x+1,\alpha}c_{x,\alpha} + c^+_{x,\alpha}c_{x+1,\alpha}\big)
+  V(n_1,\dots,n_L) + \sum_{x,a} J_x\ T_x^a  T_{x+1}^a\nonumber
\\
\phantom{H=}{}- \sum_{x,\alpha>\beta} K_x\big(
c^+_{x+1,\alpha} c^+_{x+1,\beta} c_{x,\beta} c_{x,\alpha}
+ c^+_{x,\alpha} c^+_{x,\beta} c_{x+1,\beta} c_{x+1,\alpha}\big),\label{H}
\end{gather}
where the open boundary conditions $\left(\sum_x=\sum_{x=1}^{L-1}\right)$ are supposed.
The coef\/f\/icients $t_x$, $J_x$ and~$K_x$ are positive and dependent on the site~$x$.
There are $N$ dif\/ferent f\/lavors of fermions, which are numbered by $\alpha$.
The fermion  creation  $c^+_{x,\alpha}$ and annihilation  $c_{x,\alpha}$ operators
obey the standard  anticommutation relations.

In the above Hamiltonian, $n_x=\sum_{\alpha}n_{x,\alpha}=\sum_{\alpha} c^+_{x,\alpha}c_{x,\alpha}$
is the fermion number at the $x$-th site.
The form of the potential $V(n_1,\dots,n_L)$ does not matter, the
only restriction is that it depends only on the local fermion numbers.
The Hubbard potential is a particular case
$V=U/2\sum_x n_x^2$.
It is equivalent, up to the total particle number,
to $U\sum_{x,\alpha>\beta}n_{x,\alpha}n_{x,\beta}$.

The third term is the Heisenberg interaction of $SU(N)$ spins (f\/lavors) given in
the Schwinger representation
\begin{equation}
\label{schwinger}
T_x^a=\sum_{\alpha,\beta}
c_{x,\alpha}^+ \mathcal{T}^a_{\alpha\beta} c_{x,\beta}, \qquad a=1,\dots,N^2-1=\dim SU(N),
\end{equation}
where $\mathcal{T}^a_{\alpha\beta}$ are the generators of $SU(N)$ Lie
algebra in the def\/ining representation, which are  orthogonal
with respect to the trace.
 Using the completeness relation for $SU(N)$ matrices
\[
\sum_a\mathcal{T}^a_{\alpha\beta}\mathcal{T}^a_{\alpha'\beta'}=
2\delta_{\alpha\beta'} \delta_{\alpha'\beta}  -\frac{2}{n} \delta_{\alpha\beta} \delta_{\alpha'\beta'},
\]
one can rewrite this term in the following form \cite{MA89}:
\begin{gather}
\sum_{a} T_x^a  T_{x+1}^a
=
2\sum_{\alpha,\beta} c^+_{x,\alpha}  c_{x,\beta} c^+_{x+1,\beta} c_{x+1,\alpha}
- \frac{2}{n} n_x n_{x+1}\nonumber
\\
\hphantom{\sum_{a} T_x^a  T_{x+1}^a}{}=
-2\sum_{\alpha,\beta} c^+_{x,\alpha}  c_{x+1,\alpha}  c^+_{x+1,\beta} c_{x,\beta}+2n_x- \frac{2}{n} n_x n_{x+1},\label{heis}
\end{gather}
The f\/irst term above just exchanges the f\/lavors between
adjacent sites \cite{AL86}. The last two terms depend only on $n_x$ and may be included in the potential.

The last term   in \eqref{H} describes the hopping of
fermion pairs.

The Hamiltonian preserves the $U(1)$ total charge, which corresponds to the total number
of particles
\begin{equation*}
n=\sum_{x} n_x=\sum_{x,\alpha}c_{x,\alpha}^+c_{x,\alpha}.
\end{equation*}
It is also invariant with respect to $SU(N)$ generators
\begin{equation*}
T^a=\sum_x T^a_x= \sum_{x,\alpha,\beta}c_{x,\alpha}^+ \mathcal{T}^a_{\alpha\beta}c_{x,\beta}.
\end{equation*}
The total symmetry group is $U(N)=SU(N)\times U(1)$.
The spin-raising, spin-lowering, and Cartan generators are given by
the upper triangular, lower triangular, and diagonal
matrixes. The corresponding basis is presented below
both in the def\/ining and multi-particle representations.
\begin{equation}
\label{raising}
\big(\mathcal{F}^{\alpha\beta}\big)_{\alpha'\beta'}=\delta^\alpha_{\alpha'}\delta^\beta_{\beta'},
\qquad
F^{\alpha\beta}=\sum_x c_{x,\alpha}^+ c_{x,\beta},
\qquad
F^{\alpha\alpha}=\sum_x n_{x,\alpha}.
\end{equation}

The current system  is a $SU(N)$ generalization of the $SU(2)$ Hamiltonian,
for which the Lieb--Mattis theorem  has been established
and proven already \cite{Amb92-2}.
For the Hubbard potential, the
f\/irst three terms of the Hamiltonian \eqref{H}
correspond to the $SU(N)$ Hubbard--Heisenberg chain
introduced in~\cite{MA89}.
The $SU(4)$ system without pair-hopping and Heisenberg terms has been proven to have  nondegenerate
ground state and gapless excitations \cite{LTML04}.

Each site has $2^N$ dif\/ferent states with fermion number varying from
zero to $N$.
According to~\eqref{schwinger} or~\eqref{raising},
the one-particle states $c^+_{\alpha}|0\rangle$ form the def\/ining
representation of $U(N)$.
Similarly, due to the anticommutation of the fermionic operators,
the multi-particle states $c^+_{\alpha_1}\cdots c^+_{\alpha_k}|0\rangle $
form the $\binom{N}{k}$-dimensional antisymmetric representation.
There are two singlets ($k=0,N$), which correspond to
the vacuum $|0\rangle $ and completely f\/illed states.

In this article we follow the standard way in order to establish and prove the generalized Lieb--Mattis
theorem for the described system \cite{LM62,H07}. First, we construct a basis, in which all
nonzero of\/f-diagonal matrix elements of the Hamiltonian are negative.
Next, we conf\/ine ourselves to the subspaces, where the Hamiltonian is connected,
and due to the Perron--Frobenius theorem, the lowest-energy state
is unique and positive. Employing the positivity, we compare this state with a simple trial state
and detect in this way  the containing multiplet.
Finally, using the representation theory of $SU(N)$ group, we express the ordering
of the lowest energy levels of dif\/ferent multiplets in terms of the dominance
order of the corresponding Young diagrams.

\section{Nonpositive basis}\label{section3}

The natural basis for the Hamiltonian \eqref{H} is formed by the f\/ixed particles. The coordinates
are given by the set $\{x^\alpha\}_{\alpha=1}^N$ consisting of $N$  subsets. Each subset
\[
\{x^\alpha\}=\{x^\alpha_1,\dots,x^\alpha_{M_\alpha} | x^\alpha_1<\dots <x^\alpha_{M_\alpha}\}
\]
describes the positions of the fermions carrying the f\/lavor $\alpha$, and $M_\alpha=\#\{x^\alpha\}$
is their number. Let  $M=\sum_\alpha M_\alpha$
be the total number of particles.
It appears that for each state a sign factor can be
chosen in order to make the nonzero of\/f-diagonal elements of the Hamiltonian
negative.

First, we observe from \eqref{H} and \eqref{heis} that the of\/f-diagonal part is built of
the nearest-neighbor hoppings $c^+_{x\pm1,\alpha}c_{x,\alpha}$ and their products.
Due to the appropriate signs of the
constants $-t_x$, $-K_x$ and $J_x$,
the positive values for all nonzero  matrix elements of these hoppings
imply  the non-positive values for the of\/f-diagonal matrix elements of the Hamiltonian.
The correct sign factor is encoded in the following arrangement of the fermion creation operators:
\begin{equation}
\label{basis}
|\{x^1\},\dots,\{x^N\}\rangle =
\prod_{\alpha=1}^N (c^+_{x^\alpha_1,\alpha}c^+_{x^\alpha_2,\alpha}\cdots c^+_{x^\alpha_{M_\alpha},\alpha}) |0\rangle,
\end{equation}
where the product is taken from the left to the right. In other words, we group together the particles
of the same f\/lavor ordering them according to their position.
Due to the Fermi--Dirac statistics, the hopping term  $c^+_{x\pm1,\alpha}c_{x,\alpha}$ acts
nontrivially on the states having one and only one $\alpha$-fermion per two adjacent states.
In that case, it produces a similar state~\eqref{basis} with $c^+_{x,\alpha}$ just replaced by $c^+_{x\pm1,\alpha}$
without any additional sign factor.
Therefore, the of\/f-diagonal matrix elements of the  Hamiltonian~\eqref{H} in the above basis are non-positive.

The basis \eqref{basis} has been used before to study the degeneracy of the ground states.
It was described by Af\/f\/leck and Lieb and used for the
construction of  non-positive basis for the antiferromagnetic $SU(N)$ Heisenberg chains~\cite{AL86}.
An  explicit expression similar to \eqref{basis} was written
for the extended $t-J$ and $SU(2)$ Hubbard models in \cite{Amb92-1,Amb92-2}, where it
was applied for the proof of the Lieb--Mattis theorem.
For multicomponent $t-J$ model with $SU(N_b)\times SU(N_f)$
symmetry, the similar basis ensures the nondegeneracy of the relative ground states~\cite{sorella96}.
For $SU(4)$ Hubbard model, it has been used for the study
of ground state and the proof of the Lieb--Schultz--Mattis theorem~\cite{LTML04}.

It is interesting to extract the sign factor, which was encoded in the particular arrangement
of fermion operators in \eqref{basis}, in case of the pure Heisenberg system.
Note that the Heisenberg interaction preserves the number of particles on each site. Therefore, it
may be restricted to smaller subspace, where each site contains only one of $N-1$ fundamental
(antisymmetric) representations of $SU(N)$. It is easy to express the basis \eqref{basis}
in terms of the usual Ising basis of the Heisenberg model. Set, for the simplicity, one fermion per site,
which corresponds to the def\/ining $N$-dimensional representation. Then the usual Ising (Potts) basis is
\begin{equation}
\label{ising}
|\alpha_1,\dots,\alpha_L\rangle =c^+_{1,\alpha_1}c^+_{2,\alpha_2}\cdots c^+_{L,\alpha_L} |0\rangle.
\end{equation}
The Heisenberg exchange \eqref{heis} acts on these states as
\[
\sum_a T^a_xT^a_{x+1}=2P_{x,x+1}-\frac{2}{n},
\]
where $P_{x,y}$ is the permutation of two sites. The states \eqref{basis} can be obtained by
an appropriate rearrangement of the states \eqref{ising}. Since the fermions of the same f\/lavor
are already ordered by coordinates in \eqref{basis}, we get:
\[
|\{x^1\},\dots,\{x^N\}\rangle = (-1)^{p_{\alpha_1\dots\alpha_L}}
|\alpha_1,\dots,\alpha_L\rangle,
\qquad
p_{\alpha_1\dots \alpha_L} =  \{\, \#(x<y) \; | \; \alpha_x>\alpha_y\, \}.
\]
Here $p_{\alpha_1\dots \alpha_L}$ is the number of inversions in the sequence.
The nonpositive basis in this form have  been used in the study of Heisenberg chains with
higher symmetries \cite{H04,Li01}.

Note that for the systems with ref\/lection symmetry, there is
another approach referred as a ref\/lection positivity.
It can be applied for more general class of systems, like frustrated spin models
\cite{Lieb89,LS99,Shen98}. The spin f\/lip on the half of the lattice is performed
in order to construct a new basis (distinct from \eqref{basis}) from the usual one.
The ground state wavefunction becomes positive in the new
basis. The method is problematic for higher rank $ SU(N)$  spins, since most of
the multiplets, including the def\/ining one, are not invariant under the ref\/lection.
As a result, one must either use the ref\/lected (conjugate)
representations for the half of the lattice, or conf\/ine itself to
self-conjugate ones, which are not the cases considered in this article.

\section{Relative ground states}\label{section4}

The Hamiltonian is invariant on any $M_\alpha$-subspace, which is made up of the basic states~\eqref{basis} with the same number of fermions of each type. It can be considered also as
a weight subspace under the $U(N)$ action.
The weight is given by  the set $\{M_\alpha\}_{\alpha=1}^N$ composed of
the eigenvalues of the diagonal generators from \eqref{raising}.
Note that according to the Fermi--Dirac statistics, the volume  of the chain must be large
enough in order to contain all particles:
\begin{equation}
\label{size}
L\ge\max_{1\le\alpha\le N} M_\alpha.
\end{equation}
This is the condition of the existence of $M_\alpha$-subspace for the $L$-size chain.

It is easy to see that any two basic states from the same subspace
are connected by the kinetic terms of the Hamiltonian.
Then, according to the Perron--Frobenius theorem~\cite{Lancaster},
\begin{itemize}\itemsep=0pt
\item
The relative ground state of the
Hamiltonian in any $M_\alpha$-subspace is unique, and its all coef\/f\/icients
in the basis \eqref{basis} are straightly positive:
\begin{equation}
\label{rgs}
\Omega_{M_1\dots M_N}=\sum_{\substack{\{x^1\},\dots,\{x^N\}\\ \#\{x^\alpha\}=M_\alpha}}
\omega_{\{x^1\}\dots\{x^N\}} |\{x^1\},\dots,\{x^N\}\rangle,
\qquad  \omega_{\{x^1\}\dots\{x^N\}} >0.
\end{equation}
\end{itemize}

The positivity can be used in order to trace the type of the $U(N)$
multiplet, which contains the above state.
From the basis  \eqref{basis}, we choose a trial state $\Psi_{M_1\dots M_N}$,
where the fermions of each f\/lavor
occupy successively the sites starting from the f\/irst one:
\begin{equation}
\label{trial}
\Psi_{M_1\dots M_N} =
\prod_{\alpha=1}^N ( c^+_{1,\alpha}c^+_{2,\alpha} \cdots c^+_{M_{\alpha},\alpha}) |0\rangle.
\end{equation}

For the sake of completeness,  below we present in terms of fermions  some aspects
of the representation theory  of the unitary group, which are essential in
the following discussions.
The irreducible representations of $U(N)$
are labeled by the Young diagrams $\yn$ with at most $N$ rows~\cite{Hamermesh}.
Every box of~$\yn$ is associated with a single particle,
and the number of boxes is equal to  the number of particles.
Symmetrize the f\/lavors over the rows, then antisymmetrize over the columns. The spatial coordinates
are kept f\/ixed during this process.
The states constructed in this way from all possible distributions of  f\/lavors along the
boxes of $\yn$  form an irreducible representation of the unitary group.
The states, where f\/lavors do not decrease along the rows from left to right
and increase along the columns from top to bottom, form the standard basis of the multiplet.
Among them there is the highest weight state, in which the $i$-th rows is f\/illed
by the particles with the f\/lavor $\alpha=i$. This fact can be easily verif\/ied acting
on it by the raising generators from~\eqref{raising}.

In case of two particles at sites $x\ne y$ the aforementioned procedure
extracts the symmetric and antisymmetric multiplets:
\def\YGbox{15}    
\def\YGrule{0.4}   
\[
(c^+_{x,\alpha}c^+_{y,\beta}+c^+_{x,\beta}c^+_{y,\alpha})|0\rangle
=
\YoungTab[-0.3][\normalsize]{ {{}^x\alpha,{}^y\beta} }
\;,
\qquad
(c^+_{x,\alpha}c^+_{y,\beta}-c^+_{x,\beta}c^+_{y,\alpha})|0\rangle
=
\YoungTab[-0.8][\normalsize]{ {{}^x\alpha}{{}^y\beta} }\;.
\]
The site index is mentioned at the upper left corner of the box.
Since we deal with the fermions, all sites on the same row must dif\/fer in order to
get a nonzero state.
For the state below, the $SU(N)$ spins in the brackets are symmetrized, then the corresponding spins
of each group are antisymmetrised giving rise to the presented Young tableau:
\begin{equation}
\label{exam}
(c^+_{x,\alpha}c^+_{y,\beta}c^+_{z,\gamma})(c^+_{x,\mu}c^+_{z,\nu})c^+_{x,\delta}|0\rangle
\quad\longrightarrow\quad
\YoungTab[-1.4][\normalsize]{ {{}^x\alpha,{}^y\beta,{}^z\gamma} {{}^x\mu,{}^z\nu } {{}^x\delta }}\;.
\end{equation}

Note that the symmetrization along a row is trivial if it contains particles of one species
like the rows of the highest weight state.
This would happen with the f\/irst row in the above tableau if $\alpha=\beta=\gamma$.
Similarly, the antisymmetrization along a column is trivial if its particles are on the same site like
the f\/irst column in~\eqref{exam}.

Consider now the trial  state $\Psi_{M_1\dots M_N}$.  Construct the Young tableau
with the row lengths given by the set $\{M_\alpha\}$, the $j$-th column containing particles from the $j$-th site.
Fill the f\/irst row by  $\alpha_1$-type fermions, where $M_{\alpha_1}$ is the largest number from the set, then
the second row  by  $\alpha_2$-type fermions, where $M_{\alpha_2}$ is the next largest number,
and so on. As was argued above,  the entire symmetrization-antisymmetrization
procedure is trivial (not needed).
Therefore, the trial state \eqref{trial} is really given  by the constructed Young tableau,
and it is a part of a multiplet described by the similar Young diagram.
Note that like the Fermi sea, the trial state is the most compact one: it occupies $M_{\alpha_1}$ sites,
and due to \eqref{size}  exists for any $M_\alpha$-subspace.
Therefore, the state \eqref{trial} belongs to the multiplet related to the
constructed Young diagram. Note that the last property is peculiar: the
common basic  state \eqref{basis}, in general, is not a part of a single multiplet.
Below are the examples of the trial states for $SU(3)$ chain:
\[
\Psi_{3,2,1} =\YoungTab[-1.4][\normalsize]{ {{}^1 1,{}^2 1,{}^3 1} {{}^1 2,{}^2 2 } {{}^1 3 }}
\;,
\qquad
\Psi_{2,3,1}=\YoungTab[-1.4][\normalsize]{ {{}^1 2,{}^2 2,{}^3 2} {{}^1 1,{}^2 1 } {{}^1 3 }}
\;,
\qquad
\Psi_{2,0,4}=\YoungTab[-0.8][\normalsize]{ {{}^1 3,{}^2 3,{}^3 3, {}^4 3 } {{}^1 1,{}^2 1 } }
\;.
\]
In the f\/irst state, the f\/irst site of the chain is f\/illed completely making up a singlet,
the second site has two fermions with $\alpha=1,2$, and the third one is occupied by one $\alpha=1$ fermion.
The remaining sites are empty.

Due to $U(N)$ symmetry and orthogonality of nonequivalent representations,
the projections of
$\Omega_{M_1\dots M_N}$ on dif\/ferent sectors corresponding to the nonequivalent
multiplets produce orthogonal states with the same energy.
Due to the uniqueness of the relative ground state, it must belong to only one sector.
According to \eqref{rgs}, the state $\Psi_{M_1\dots M_N}$ being a basic state
participates in the decomposition of the relative ground state $\Omega_{M_1\dots M_N}$
overlapping it. So, both states are the members of equivalent multiplets.
We conclude that
\begin{itemize}\itemsep=0pt
\item
The relative ground state state in $M_\alpha$-subspace
belongs to a single irreducible $U(N)$ representation
characterized by the Young diagram $\yn_{M_\alpha}$
with row lengths given by the nonzero numbers from the set $\{M_\alpha\}$.
\end{itemize}

\section{Ordering of energy levels}\label{section5}

Due to $U(N)$ symmetry, the Hamiltonian of the extended Hubbard chain~\eqref{H}
remains invariant on the individual sectors  combining the equivalent
representations. These sectors are labeled by the Young diagrams.
The number of boxes is the quantum number of the $U(1)$ subgroup and corresponds
to the total number of particles.

Denote by $E(\yn)$ the lowest energy level among
all  multiplets of the same equivalence class~$\yn$.
In fact, the relative ground state $\Omega_{M_1\dots M_N}$ has the lowest energy
level $E(\yn_{M_\alpha})$ because any~$\yn_{M_\alpha}$ multiplet has a representative
in the corresponding $M_\alpha$-subspace. The nondegeneracy of the level  $E(\yn_{M_\alpha})$
follows directly form the uniqueness of the relative ground state.

Note that the relative ground states in all $M'_{\alpha}$-subspaces obtained by
rearrangements of the set $\{M_\alpha\}$ are related to the same Young diagram.
In fact, all they are members of the same~$\yn_{M_\alpha}$ multiplet \cite{H04}.
This fact ref\/lects the discrete symmetry of the Hamiltonian  with  respect to
the permutation group of $N$ f\/lavors, which is a discrete subgroup of the unitary group.
Therefore, one can consider without loosing the generality, the $M_\alpha$-subspaces
with nonascending sequences $M_1\ge\dots\ge M_N$.
The corresponding relative ground states are the highest weight states of the
lowest energy $\yn_{M_\alpha}$ multiplet.

Consider now two dif\/ferent Young diagrams $\yn_{M_\alpha}$ and $\yn_{M'_\alpha}$
and try to compare the related minimal energies.
Suppose that  the highest weight vector of the f\/irst multiplet is also a weight (evidently, not
highest) of the second one.
This means that the irreducible representation generated by the relative
ground state  $\Omega_{M'_1\dots M'_N}$ has also a representative in the
$M_\alpha$-subspace. Of course, both states dif\/fer. Then, due to
the uniqueness of the relative ground state,  this representative, together
with the whole multiplet, has a higher energy  than the $\Omega_{M_1\dots M_N}$ has,
i.e.\
$E(\yn_{M'_\alpha})>E(\yn_{M_\alpha})$.

This relation introduces some ordering among the representations,
which can be formulated in an elegant way in terms
of  Young diagrams. Namely, in this case $\yn_{M_\alpha}$ may be obtained from~$\yn_{M'_\alpha}$ by the displacement of some of its boxes from the upper rows to the lower ones,
which we note shortly as $\yn_{M'_\alpha}\succ\yn_{M_\alpha}$ \cite{H04}.
In the representation theory of the symmetric group, this is known as a dominance
order~\cite{Mac}.
For example, for unitary group with $N\ge 4$ we have:
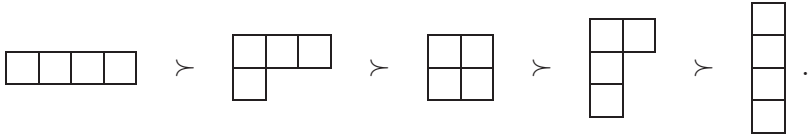
\[
{\yc
\young(\mb\mb\mb\mb) \quad \succ \quad
\young(\mb\mb\mb,\mb) \quad \succ \quad
\young(\mb\mb,\mb\mb)\quad \succ \quad
\young(\mb\mb,\mb,\mb) \quad \succ \quad
\young(\mb,\mb,\mb,\mb)
\;\; .
}
\]

The dominance order is a partial one. There are Young diagrams, which are
not related by this order for higher ($N>2$) algebras and higher ($M>5$) box numbers,
like the following ones:
{\tiny $\yc \yng(4,1,1)\;$} and {\tiny $\yc \yng(3,3)\;$}.
The  Young diagrams with dif\/ferent number of boxes are not related to each other also.

In summary, we have proved that for the extended Hubbard model \eqref{H},
\begin{itemize}\itemsep=0pt
\item
The minimum energy levels in the sectors characterized by dif\/ferent
Young diagrams satisfy the ordering rule
\begin{equation}
\label{lm}
E(\yn_2)>E(\yn_1)\qquad \text{if} \quad \yn_2\succ \yn_1;
\end{equation}
\item
The levels $E(\yn)$ are nondegenerate, up to the trivial $SU(N)$ degeneracy.
\end{itemize}

These results are in agreement with those obtained for the $SU(2)$ system by Xiang
and d'Ambrumenil~\cite{Amb92-2}. In that case, the  Young diagram is labeled by
the spin quantum number~$S$, and  the usual Lieb--Mattis ordering rule
$E(S_2)>E(S_1)$ if $S_2> S_1$  is fulf\/illed~\cite{LM62}.
For the pure Heisenberg system in the def\/ining representation, the system is reduced
to the Sutherland chain~\cite{Suth}. A similar ordering rule
for that system has been formulated and proven in~\cite{H04}.

The described ordering was used  already in the one-dimensional
many-particle quantum mechanics by Lieb and Mattis in order to compare
the minimum energies of the  wavefunctions with dif\/ferent
permutation symmetries~\cite{LM62'}.
This is not surprising, because these symmetry classes are also described by the Young diagrams.
The pouring principle obtained in \cite{LM62'} is just the ``reverse'' version of the
energy level ordering obtained above (see~\eqref{lm'} below).
We show in the last section that  for the particles with $SU(N)$ spin degrees of freedom,
the spatial and spin parts of the fermionic wavefunction
are described by conjugate Young diagrams. This leads to the ``direct'' ordering
for the spins in agreement with our results above.

\section{Ground state}\label{section6}


Although the dominance order is partial, there is a lowest diagram $\yn_\text{gs}$
among all diagrams containing the same amount  of boxes: $\yn\succ\yn_\text{gs}$.
All columns in~$\yn_\text{gs}$  have the maximal length $N$ besides the last one having
\begin{equation}
\label{m}
m=M\mod{N}
\end{equation}
boxes, where $M$ is the total box number.    $\yn_\text{gs}$   corresponds to the $m$-th order
antisymmetric representation.
According to the ordering rule \eqref{lm}, the sector def\/ined by  $\yn_\text{gs}$   has
the lowest energy value amon other sectors:  $\yn_\text{gs}$. So,
\begin{itemize}\itemsep=0pt
\item
The ground states of the extended Hubbard chain \eqref{H} with $M$ particles form a unique
$\binom{N}{m}$-dimensional antisymmetric $SU(N)$ multiplet.
\item
In particular, if
the number of particles is the multiplicity of $N$, the ground state is a~unique
singlet.
\end{itemize}

For $SU(2)$ case, the ground state is a  spin singlet for even number of particles, while for odd
number it is a spin doublet in agreements with the results \cite{Amb92-2}.
For  $SU(3)$ symmetric chain, depending on the value of the remainder \eqref{m}, the ground state
is a singlet, a three-dimensional def\/ining representation $\mathbf{3}$, or its complex
conjugate one  $\mathbf{\bar{3}}$. They are presented below in case of six, seven,
and eight particles.
\begin{equation}
\label{su3}
{\yc
\yn_\text{gs}^\mathbf{0}=\yng(2,2,2)\;,
\qquad  \yn_\text{gs}^\mathbf{3}=\yng(3,2,2)\;,
\qquad  \yn_\text{gs}^\mathbf{\bar{3}}=\yng(3,3,2)\;.
}
\end{equation}
Note that using the described method we can not compare the ground states having
dif\/ferent  amount of particles.

Consider now the solvable free fermion case  when only the hopping
term survives in the Hamiltonian~\eqref{H}, and the ground state has a
very simple form.
The digitalization of the bilinear Hamiltonian is reduced to the digitalization
of $L\times L$ matrix composed from the coef\/f\/icients~$t_x$. There are $L$ energy
eigenvalues~$\varepsilon_k$, which we arrange is ascending order: $\varepsilon_1<\dots<\varepsilon_L$.
Under the periodic boundary conditions and translation invariance,
they are reduced to  $\varepsilon_k=4t\sin^2(\pi (k-1)/L)$. Note that here there is the twofold degeneracy
$\varepsilon_k=\varepsilon_{L-k}$ due to the ref\/lection invariance.
This degeneracy is removed in the general case, but the degeneracy on the f\/lavor quantum number
still remains.  In the ground state with $M$-fermion, all lowest levels $\varepsilon_k$ are completely f\/illed
by $N$ fermions of dif\/ferent f\/lavors up to the Fermi level $k_F$, which is f\/illed
partially by $m=M-k_F N$ particles. The fermions with the completely f\/illed levels form, of course,
singlets, while  the remaining fermions form $m$-th order antisymmetric multiplet on the Fermi level.
This is exactly the same picture as we obtained for the interacting system.
In the examples considered in~\eqref{su3}, the
ground states for noninteracting system are described by the following Young tableaux:
\[
\Omega_\text{gs}^\mathbf{0}=\YoungTab[-1.4][\normalsize]{ {{}^1 1,{}^2 1} {{}^1 2,{}^2 2 }   {{}^1 3, {}^2 3 }}
\;,\qquad
\left(\Omega^\mathbf{3}_\text{gs}\right)_\alpha
=\YoungTab[-1.4][\normalsize]{ {{}^1 1,{}^2 1,{}^3 \alpha} {{}^1 2,{}^2 2 } {{}^1 3, {}^2 3 }}
\;,\qquad
\big(\Omega^\mathbf{\bar{3}}_\text{gs} \big)_{\alpha\beta}
=\YoungTab[-1.4][\normalsize]{ {{}^1 1,{}^2 1,{}^3 \alpha} {{}^1 2,{}^2 2, {}^3 \beta } {{}^1 3, {}^2 3 }}
\;.
\]
Now the number at the upper left corners is the energy quantum number $k$, but not the
space position as before. The $\alpha$, $\beta$ are $SU(3)$ f\/lavors which label the three states
of the multiplet.

The relative ground states
$\Omega_{M_1\dots M_N}$ can be constructed
in the same way. First choose the largest number $M_\alpha$ from the set and f\/ill the f\/irst
level by $M_\alpha$ fermions of f\/lavor $\alpha$, then choose the next largest number
$M_\beta$   and f\/ill the second
level by $M_\beta$ fermions of f\/lavor $\beta$, and so on. This state is similar to the
trial state \eqref{trial} used to detect the type of the representation of  $\Omega_{M_1\dots M_N}$
for the interacting system. The only dif\/ference is that again, instead of coordinates
one must use the energy quantum number. The $SU(N)$ structure of these tree states
is the same.  Moreover, the relative ground state of noninteracting system can be used
as a trial state instead of \eqref{trial} as it was done for $SU(2)$ Hubbard model \cite{LM62',Amb92-2}.
Indeed, it overlaps with the interacting state  since
both states are positive in the basis \eqref{basis} (see \eqref{rgs}).

\section[Energy level ordering for the quantum mechanical system with
interacting $SU(N)$ fermions]{Energy level ordering for the quantum mechanical system\\ with
interacting $\boldsymbol{SU(N)}$ fermions}\label{section7}

Consider the  quantum mechanical system of one-dimensional identical
fermions with $SU(N)$ internal degrees of freedom and the interaction
depending on the spatial coordinates only:
\begin{gather}
\label{QM}
H_\text{QM}=-\frac{1}{2m}\sum_{i=1}^M \frac{\partial^2}{\partial x_i^2} + V(x_1,\dots,x_M).
\end{gather}
The potential is invariant under the particle exchange and must be integrable.
For the  $SU(2)$ system,  Lieb and Mattis proved the antiferromagnetic ordering
 of the energy levels \cite{LM62'} (see also~\cite{Mattis}).
In this section we apply their result to higher unitary symmetries
and  discuss the relation with the results obtained in the previous sections.

For the distinguishable particles, the  spatial and spin degrees of freedom are decoupled. The
stationary states are the products of the spatial and spin parts, and the spectrum is determined by
the former.
Due to the exchange invariance of the Hamiltonian $H_\text{QM}$, the spatial eigenfunctions
are classif\/ied by their symmetries with respect to the permutations
$\mathcal{S}_M^\text{space}$  of the spatial coordinates.
According to the representation theory of the symmetric group, the symmetry classes are described
by Young  diagrams $\yn'$ \cite{LM62',Landau,Hamermesh}. For a given distribution of the coordinates along the
boxes, it  def\/ines a similar symmetrization-antisymmetrization procedure as for~$SU(N)$ case.
The def\/ined map is a projector, and the projectors
constructed from the dif\/fe\-rent Young diagrams are mutually orthogonal.
The function $\phi_{\yn'}(x_1,\dots,x_M)$ has the  symmetry class~$\yn'$,
if the associated projector does not change it.
It generates an irreducible representation of the symmetric group.
The standard basic states correspond to the distributions, in which the indexes
increase along the rows from left to right and along the columns from top to bottom.
The totally symmetric and antisymmetric representations are one-dimensional
and are described by one-row and one-column Young diagram respectively, while the
others have higher dimensions.

The function $\phi_{\yn'}(x_1,\dots,x_M)$ is separately antisymmetric in the variables
related to the same column and satisf\/ies the following equations:
\begin{equation}
\label{sym-cond}
\left[1-\sum_{j:\;\text{col}(j)=c }P_{ij} \right] \phi_{\yn'}(x_1,\dots,x_M)=0
 \qquad
 \text{for any $i$ with $\text{col}(i)>c$}.
\end{equation}
Here  $\text{col}(i)$ is the index of the column in $\yn'$ containing $x_i$, these are counted
from left to right. $P_{ij}$ permutes $x_i$ and $x_j$.
These relations mean that the antisymmetrization of the column variables with any variable
located on the right hand side must vanish, since it
has been  already symmetrized with a variable from that column in the
Young projector.

Denote now by $E(\yn')$ the lowest energy level among the  states, which belong to the
symmetry class def\/ined by $\yn'$.  These levels are nondegenerate (up to the coordinate
permutations) and obey the following
ordering rule \cite{LM62'}:
\begin{equation}
\label{lm'}
E(\yn'_1)>E(\yn'_2) \qquad \text{if}\quad \yn'_1\prec \yn'_2.
\end{equation}
 In particular, the highest energy level
is totally antisymmetric in spatial coordinates,  while the lowest level is
totally  symmetric. Here and in the following, the notation $\phi_{\yn'}$  will
be used  for the lowest-energy state with  $\yn'$ symmetry.

For indistinguishable fermions, according to the Pauli exclusion principle, the
wavefunction of the entire system must be antisymmetric under the interchange of
individual particles. This can be achieved by the selection of appropriate
spin wavefunctions. For the  symmetric spatial part, the spin part must
me antisymmetric, and vice versa. For more general symmetry classes,
the entire wavefunction is ``entangled'', i.e.\ a superposition
of the products of spin and spatial parts. The general construction
can be f\/igured out using the representation theory.

The Hamiltonian \eqref{QM} has the trivial unitary $SU(N)$  and  permutation
$\mathcal{S}_M^\text{spin}$
symmetries, which act on the spin variables and are mutually independent.
The representations of both groups are classif\/ied by the Young diagrams $\yn$
with with $M$ boxes and at most $N$ rows.
According to the Schur--Weyl duality, their joint action
 decomposes into a direct sum of tensor products of irreducible modules:
\[
 \sum_\yn \pi_\yn \otimes \rho_\yn.
\]
Here  $\pi_\yn$  and $\rho_\yn$ are the irreducible
representations of $\mathcal{S}^\text{spin}_M$ and $SU(N)$ correspondingly.
This property implies, in particular, that the highest weight vectors of all
$SU(N)$ multiplets of type~$\rho_\yn$ form an irreducible
$\mathcal{S}^\text{spin}_M$-representation
of type $\pi_\yn$.
Similarly, the action of the common symmetry group
$\mathcal{S}_M^\text{space}\times \mathcal{S}^\text{spin}_M\times SU(N)$
on the total space of the states decomposes as
\begin{equation}
\label{schur-weyl}
\sum_{\yn',\yn}  \pi_{\yn'}\otimes\pi_\yn \otimes \rho_\yn.
\end{equation}
Here $\pi_{\yn'}$ the (unique) irreducible representation
of the symmetric group formed by the lowest-energy
spatial wavefunctions with the symmetry class $\yn'$.
The physical space of states in our case
is the subspace of \eqref{schur-weyl} formed by the antisymmetric wavefunctions.
It forms the antisymmetric representation
of the  symmetric group corresponding to the image of the  injective diagonal homomorphism
$\mathcal{S}_M\rightarrowtail \mathcal{S}^\text{space}_M\times\mathcal{S}^\text{spin}_M$.
The representation $\pi_{\yn'}\otimes\pi_\yn$ can be treated as a tensor
product (or Kronecker product) of two $\mathcal{S}_M$ modules.
According to the representation theory of the symmetric group, the antisymmetric representation
appears only in the tensor product of two conjugate representations, while
the symmetric representation appears in the products of~two equivalent ones \cite{LM62',Hamermesh}.
Their  multiplicities are equal to  one.
The conjugate Young diagrams~$\yn$ and $\tilde\yn$  are ref\/lections of each other
with respect to the main diagonal, i.e.\ the rows (columns) of~$\yn$
are replaced by columns (rows) of~$\tilde\yn$.
Both $\pi_\yn$ and $\pi_{\tilde\yn}$
have the same dimension, and one can obtained from another by multiplication on the
antisymmetric representation.

So, only the terms with $\yn'=\tilde\yn$ are relevant for the fermionic states
in the sum \eqref{schur-weyl}.
Extracting them and applying  the formal Clebsch--Gordan series, we
arrive at
\begin{equation*}
\sum_{\yn}  \pi_{\tilde\yn}\otimes\pi_\yn \otimes \rho_\yn=
\sum_\yn\left[\pi_\text{asym}^{\tilde\yn\times\yn}\otimes\rho_\yn+
\sum_{\yn'\ne\text{asym}}
\pi_{\yn'}^{\tilde\yn\times\yn}\otimes\rho_\yn\right].
\end{equation*}
Thus, the fermionic  wavefunctions constructed from $\phi_{\tilde\yn}$
 form a unique $\rho_\yn$-type $SU(N)$ multiplet.
Its highest weight state $\Phi^{\tilde\yn\times\yn}_\text{asym}$ has a simple ``canonical'' form,
which was constructed and applied in case of $SU(2)$ fermions in \cite{LM62'}.

Let the $M_i$ be the length of the $i$-th row of $\yn$.
Suppose, for certainty, that the f\/irst $M_1$
variables of $\phi_{\tilde\yn}(x_1,\dots, x_M)$ are positioned
on the f\/irst column in the downward direction,
the second $M_2$ ones are on the second column, and so on.
Then we have (up to a normalization factor):
\begin{equation}
\label{wavefunction}
\Phi^{\tilde\yn\times\yn}_\text{asym}=\sum_{P\in \mathcal{S}_M}(-1)^P \pi_{\tilde\yn}(P)\phi_{\tilde\yn}(x_1,\dots, x_M)
\pi_\yn(P)|\underbrace{1\dots1}_{M_1}\,\underbrace{2\dots2}_{M_2}\ldots \ldots \underbrace{N\dots N}_{M_N}\rangle.
\end{equation}
It belongs to the Kronecker product $\pi_{\tilde\yn}\otimes\pi_\yn$ and does not vanish,
since the coef\/f\/icient in front of the ordered spin
state  is proportional to $\phi_{\tilde\yn}(x_1,\dots, x_M)$.
Due to the factor $(-1)^P$ (the parity of $P$),   the above state is totally
antisymmetric under an interchange of two particles.
Finally, its  $SU(N)$ weight coincides with the highest weight of $\rho_\yn$.
From the uniqueness condition, we conclude that the expression \eqref{wavefunction} is correct.

The  fact that the wavefunction \eqref{wavefunction} is the highest weight state,
i.e.\ the spin-raising generators (given in \eqref{raising}, $\alpha<\beta$)
annihilate it, can be verif\/ied  independently by direct calculations.
Below we demonstrate this for $F^{12}$, the other generators can be handled in the same way.
\begin{gather}
\rho_\yn\big(F^{12}\big)\Phi^{\tilde\yn\times\yn}_\text{asym}
=\!\!\sum_{P\in \mathcal{S}_M}(-1)^P P\phi_{\tilde\yn}(x_1,\dots, x_M)
\sum_{i=M_1+1}^{M_2}P\mathcal{F}^{12}_i|\underbrace{1\dots1}_{M_1}\,\underbrace{2\dots2}_{M_2}\ldots  \rangle
\nonumber\\
 \phantom{\rho_\yn(F^{12})\Phi^{\tilde\yn\times\yn}_\text{asym}}{}
 =\!\!\sum_{P\in \mathcal{S}_M}(-1)^P P\phi_{\tilde\yn}(x_1,\dots, x_M)
P\left[1+\!\sum_{i=M_1+2}^{M_2}\!P_{M_1+1\,i}\right]|\underbrace{1\dots1}_{M_1+1}\,\underbrace{2\dots2}_{M_2-1}\ldots  \rangle\!\!\!\!
\label{E-Phi}\\
\phantom{\rho_\yn(F^{12})\Phi^{\tilde\yn\times\yn}_\text{asym}}{}
=\!\!\sum_{P\in \mathcal{S}_M}(-1)^P P\left[1-\!\sum_{i=M_1+2}^{M_2}\!P_{M_1+1\,i}\right]\phi_{\tilde\yn}(x_1,\dots, x_M)
P|\underbrace{1\dots1}_{M_1+1}\,\underbrace{2\dots2}_{M_2-1}\ldots  \rangle=0.\!\nonumber
\end{gather}
Here the matrix $\mathcal{F}^{12}_i$ acts on $i$-th spin.
In the f\/irst equation we have used the commutativity of unitary and symmetric groups.
The second equation employs the def\/inition of $\mathcal{F}^{12}$ in \eqref{raising}.
In the third equation we have changed the summation index  $\left[\sum_P\to\sum_{PP_{M_1+1\,i}}\right]$, which
alters also the sing in the square brackets since $(-1)^{PP_{M_1+1\,i}}=-(-1)^P$. The last equation
in~\eqref{E-Phi} is the consequence of the relations~\eqref{sym-cond}.

So, we come to the conclusion that the spin and spatial parts of the fermionic wavefunction are described
by the conjugate Young diagrams.
It is clear that the conjugation inverts the dominance order: if $\yn_1\succ\yn_2$ then
$\tilde\yn_1\prec\tilde\yn_2$ and
vice versa \cite{LM62'}. Therefore, in terms of the  $SU(N)$ representations,
the ``reverse'' ordering rule \eqref{lm'} changes to the ``direct'' one, which corresponds
to the antiferromagnetic ordering \eqref{lm} established for $SU(N)$ Hubbard chain
in the previous sections. For the Sutherland chain the similar ordering has been
established in  \cite{H04}.
Recall that the spin Young diagrams have no more than $N$ rows, hence the spatial ones must
have no more than $N$ columns. For usual $SU(2)$ spin they correspond to two-row and two-column
diagrams respectively \cite{LM62'}.

Note that for the indistinguishable  bosons, the reverse ordering rules \eqref{lm'} takes place.
The total wavefunctionnow must be symmetric, which must be composed from
the equivalent representations of the spatial and spin symmetric groups
(must apply $\yn'=\yn$  in  \eqref{schur-weyl},
replace $\tilde\yn\to\yn$ and omit the parity factor in  \eqref{wavefunction}).

Finally, we mention  that although the ordering of energy levels for both lattice \eqref{H} and quantum mechanical~\eqref{QM} systems are similar, there is an essential dif\/ference between these
two models. The second system has permutation symmetry with respect to the coordinate
exchange, which leads to the separation of the spin and spatial degrees of freedom.
The second system does not possess the spacial symmetry at all (despite of possible
ref\/lection invariance in case of the appropriate choice of constants),
it has spin interactions, and  the
lowest-energy state \eqref{rgs} can not presented in a factorized form like~\eqref{wavefunction}.

\subsection*{Acknowledgements}
The author is grateful to E.~Chubaryan and R.~Avagyan for simulating discussions.
This work was supported by the grants UC-06/07, ANSEF 2229-PS, and Volkswagen
Foundation of Germany.

\pdfbookmark[1]{References}{ref}
\LastPageEnding

\end{document}